# Self-foaming, Sintering-resistant Iron-Tungsten Powders Enable High-Cycle Thermochemical Hydrogen Storage

Jie Qi*[1], David C. Dunand*

Department of Materials Science and Engineering, Northwestern University, Evanston, IL 60208, USA

[1]currently at: Thrust of Sustainable Energy and Environment, The Hong Kong University of Science and Technology (Guangzhou), Guangzhou, Guangdong, China

* Corresponding authors: dunand@northwestern.edu; Phone number: +1-847-491-5370

jieqi@hkust-gz.edu.cn; Phone number: +1-434-327-0269

**Abstract**

$H_2$-$H_2O$ redox cycling of iron powder beds at 650-800 °C offers a compact, safe, economical hydrogen storage method, but sintering-induced capacity loss has stalled its scalability for decades. Here, we show that adding redox-active tungsten to Fe powders solves this problem in static powder beds: Fe-25W (at%) alloyed powder self-foams during redox cycling via W gas-phase transport, increasing porosity and preserving capacity. In a custom automated reactor, a kilogram-scale powder bed reversibly stores 42g $H_2$ and sustains 96±3% capacity utilization over 30 redox cycles. Temperature-resolved *in-situ* X-ray diffraction reveals a chemical-vapor-transport-mediated self-foaming mechanism that redistributes W to refine the microstructure, complemented by a contact-barrier stabilization mechanism during high-temperature holds. Partial-capacity cycling up to 90 cycles further confirms sintering resistance under incomplete redox conditions. These results establish Fe-W powder beds as a robust, scalable, and compact platform for safe, stationary hydrogen storage.

**Introduction**

Hydrogen plays a pivotal role in the chemical industry, underpinning essential sectors such as ammonia and methanol production, fuel refining, green steelmaking, and sustainable aviation fuel. These stationary applications currently account for > 90% of hydrogen end use [1]. Yet, the deployment of hydrogen, particularly green hydrogen produced intermittently during windows of low-cost power, remains constrained by storage and transportation which are the indispensable bridge between upstream production and downstream end use (Figure 1a). This bottleneck has persisted for decades because hydrogen has intrinsically low volumetric density, and the high-pressure storage of a flammable, highly diffusible gas imposes substantial safety and engineering burdens. Consequently, incumbent storage solutions such as compression, cryogenic liquefaction, and chemical or solid-state carriers incur major penalties in volumetric energy density, energy consumption, cost, lifetime, and infrastructure complexity[2], limiting the extent to which advances in hydrogen (especially green hydrogen) production can translate into deployable energy and chemical systems.

Among the many energy/hydrogen carriers, selecting a scalable hydrogen storage medium reduces to a small set of metrics (Figure 1b): volumetric and gravimetric energy density, cost, and crucially, the practicality of carrier regeneration over repeated storage-release cycles. These considerations motivate attention to abundant metals such as iron[3], which are standing out as attractive thermochemical energy carriers (Figure 1b). In the $H_2$-$H_2O$ redox cycle (Figure 1d), hydrogen reduces iron oxide to metallic iron while producing water, effectively translating hydrogen's chemical potential into stable iron that can be re-oxidized with steam to release $H_2$ on demand[3,4]. The highly reactive hydrogen atoms are safely stored into water. Iron, therefore, offers a rare combination of abundance, reversibility, safety, and high volumetric storage potential, positioning it as a uniquely scalable platform for thermochemical hydrogen storage, provided its long-standing durability limitations can be overcome.

While fluidized bed systems using various oxygen carriers can undergo numerous redox cycles, fixed powder bed configurations show much higher sintering tendencies, lack of durability and cyclability[5,6]. The large cyclical volumetric changes during high-temperature redox $H_2$-$H_2O$ cycling drive sintering and coarsening of iron powders packed in static beds, reducing reactive surface area and closing gas-transport pathways, thereby slowing reaction kinetics and reducing usable capacity[5]. Prior sintering inhibition strategies have achieved only partial success. Architected porous morphologies produced by forming routes (for example, freeze casting) can delay densification and extend lifetime, yet coarsening under repeated high-temperature redox conditions is still occurring [7,8]. Alternatively, inert "skeleton" additives, metals such as Ni[7,9] or Cu[10] that are not oxidized by steam, or stable oxides such as $CeO_2$, or $ZrO_2$[11] that are not reducible by hydrogen (as indicated by the Ellingham diagram[12] in Figure 1c), can limit iron-iron contact, yet they typically delay rather than eliminate sintering at high cycle numbers and impose a dead-mass penalty in energy density and system cost. Recently, adding redox-active additions to iron, such as molybdenum and tungsten (Figure 1c),[13–15] has led to promising sintering resistance during high-cycle $H_2$-$H_2O$ redox reactions in static powder beds.

Here, we establish iron-tungsten powders as a promising and scalable static powder-bed platform for hydrogen storage, systematically examining their storage-release performance, porosity-preserving mechanisms, the impact of storage temperature, and the effects of incomplete storage on powder porosity and lifetime. Results show that tungsten remains redox-active and the bed undergoes a self-foaming evolution during cycling that resists sintering and even refines the microstructure, expanding accessible reaction interfaces. Crucially, this system is demonstrated beyond gram-scale tests: we operate kilogram-scale powder inventories (1.5 kg, a ~1000× scaled up from typical laboratory studies) to validate performance under scale-relevant conditions. Results show that, even at the kg scale, gravity-driven sintering[16] in the powder bed can be avoided by the iron-tungsten powders' self-foaming effect, further validating their suitability for industrial deployment in static beds. As positioned against incumbent storage routes (Figure 1e), Fe-W powder beds combine intrinsic safety (no flammable, explosive, high-pressure gas), high volumetric storage density (13.6 MJ/L, ~3× that of compressed $H_2$), and low system capital expenditures. Despite incorporating high-density W, the gravimetric energy density remains comparable to most alternative storage options, while remaining below that of the pricy liquid ammonia approach (Figure 1e). These attributes, together with a credible pathway to long cycle life, make Fe-W particularly well matched to stationary hydrogen storage (which accounts for >90% of end-use applications[1]) in chemicals, refining, and steel production, where footprint and reliability dominate over gravimetric constraints.

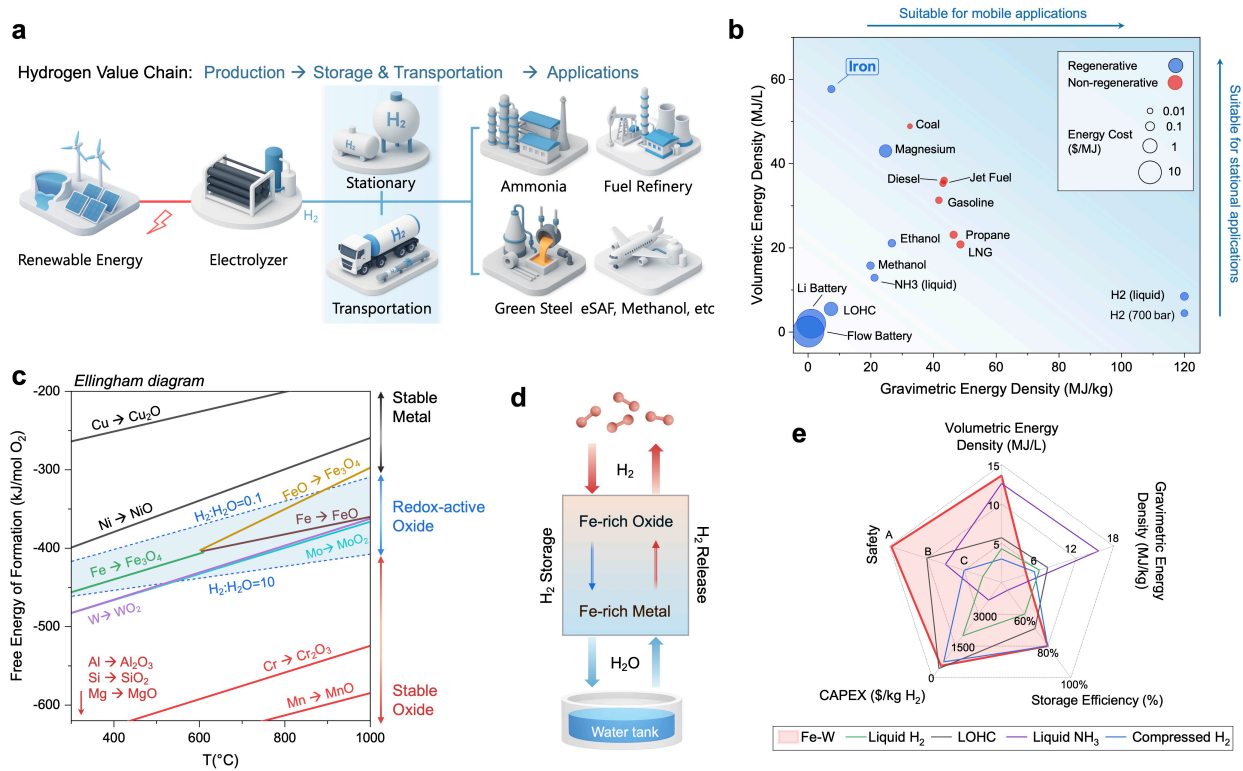

Figure 1: **Iron-based thermochemical carriers for hydrogen storage**. **a**, Schematic of the hydrogen value chain spanning production, storage/transport, and end-use sectors. **b**, Comparison of representative energy carriers by gravimetric (x axis) and volumetric (y axis) energy density; symbol size denotes energy cost, and color indicates regenerative (blue) versus non-regenerative (red) carriers. **c**, Ellingham diagram showing the standard Gibbs free energy of oxide (and water) formation versus temperature, indicating thermodynamic stability and redox feasibility. The redox-active region (blue shading) is bounded by $H_2$:$H_2O$ ratios of 0.1 and 10 (blue dashed lines). Materials above this window are considered as oxidation-stable metals, whereas those below are reduction-stable oxides. **d**, Conceptual schematic of iron-based hydrogen storage via the $H_2$-$H_2O$ redox cycle. **e**, Technology differentiation map comparing the Fe-W system developed here with incumbent industrial hydrogen storage methods in terms of volumetric and gravimetric energy density, safety, capital expenditure (CAPEX), and storage efficiency. Fe-W offers a favorable combination of safety, volumetric density, and projected cost, with gravimetric density as the primary trade-off, positioning

it for stationary storage applications. LOHC refers to liquid organic hydrogen carrier. The compressed hydrogen case uses 700 bar as the pressure. Values used in sub-figures b and e are given in Table S1-2.

**Main**

**Hydrogen storage cycling without capacity degradation**

An Fe-25W (at%) loose powder bed, prepared by hydrogen reduction at 800 °C of a blend of $Fe_2O_3$ and $WO_3$ powders, was subjected to alternating hydrogen and steam flows in a custom-designed, automated reactor (Figure 2a), leading to the following redox reaction:

$$3Fe + 3Fe_2W + 20H_2O \leftrightarrow 3Fe_3O_4 + 2FeWO_4 + 20H_2 \qquad [1]$$

The powder bed mass was 1.5 kg in the oxidized state ($Fe_3O_4$ + $FeWO_4$) and 1.1 kg in the reduced state (Fe + $Fe_2W$), corresponding to a storage capacity of 42.4 g $H_2$ (472 L at standard temperature and pressure [STP: 0 °C and 1 bar]) and an equivalent storage mass fraction of 3.8 wt% (0.0424/1.1). Unless noted otherwise, redox cycling was conducted at 800 °C. Thirty storage-release redox cycles were completed (Figure 2b). The first five were deliberately partial to validate the system's automated operation and its safe, unattended cycling under high $H_2$ flow. The subsequent 25 cycles were performed aiming for complete conversion: the Fe-W bed sustained high capacity utilization throughout, averaging 96 ± 3%, and retained a highly porous powder morphology after 30 cycles, as shown in the Scanning Electron Microscopy (SEM) micrograph in Figure 2b. In contrast, a W-free iron powder bed subjected to the same protocol exhibited rapid performance degradation: capacity utilization fell to 11 ± 1% as the powder sintered into a dense bulk (bottom micrograph in Figure 2b), severely limiting hydrogen/stream transport and reducing usable capacity.

Hydrogen storage and release processes remained highly reproducible over 30 cycles, with representative cycles 10, 20, and 30 exhibiting highly similar rate and capacity profiles (Figure 2c). Oxidation (hydrogen release) proceeds substantially faster than reduction (hydrogen storage), showing a smooth, monotonic rate decay, whereas reduction proceeds in two kinetic regimes: an initial fast stage (~5 h) followed by a slower stage (~15 h). This persistent kinetic asymmetry and the two-stage reduction, maintained without drift over repeated redox cycles, points to a stable, phase-governed reaction sequence that we describe mechanistically in the following section.

Phase evolution during cycling (Eqn.[1]) remains remarkably stable under repeated operation. Ex-situ X-ray diffraction (XRD) patterns collected in the fully oxidized (hydrogen released) and fully reduced (hydrogen stored) states are consistent across cycles 10, 20, and 30 (Figure 2d), indicating robust, reversible phase chemistry at scale during repeated operation. In the reduced state, the presence of broadened diffraction peaks, attributable to nanocrystalline $Fe_2W$ phase, indicates the formation of fine microstructure within the cycled powder bed, which is corroborated by SEM micrographs (Figure 2e). The oxidized material is generally coarser than the reduced material, consistent with volume and phase changes associated with oxidation. The dominant long-term bed morphology change trend is counterintuitive: the powder bed undergoes a pronounced self-foaming process, evolving toward microstructural refinement rather than densification and sintering. This self-generated porosity preserves gas-transport pathways and expands the accessible surface area, thereby enlarging the effective gas-solid reaction interface and rationalizing the sustained nearly full capacity utilization observed during cycling.

While the cycling was terminated after 30 cycles, the observed progressive self-foaming and microstructural refinement strongly indicate that continued cycling is unlikely to lead to the sintering-driven capacity decay typical of pure iron powders. Meanwhile, the electrolyzer-limited hydrogen supply prolonged the reduction step, making additional cycles time-intensive. Notably, removal of the fully reduced bed for ex-situ characterization after cycle 30 triggered localized exothermic reactions ("smoldering") upon air exposure (Figure S5), which was not observed after cycles 10 or 20. This observation is consistent with the finer, more porous microstructure of the highly cycled bed (high-resolution SEM images of cycle 30 powder bed are shown in Figure S2), where the porous reduced powder exhibits an increased metal-air contact area, leading to the self-sustaining oxidation reaction upon exposure to air.

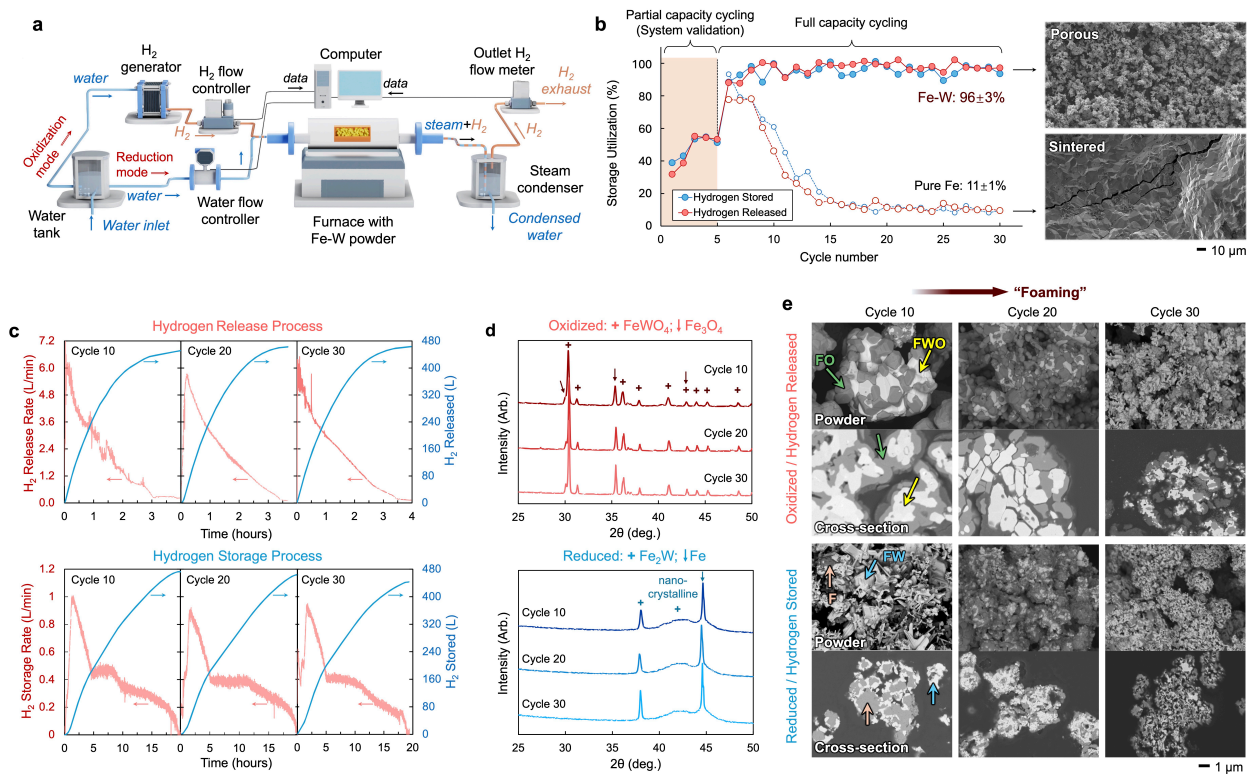

Figure 2: **Hydrogen storage cycling performance of the Fe-W system. a**, Automated cycling platform for hydrogen storage. Hydrogen (from an electrolyzer) and steam (generated by metered water injection into the furnace) are alternately supplied to the Fe-W powder bed. The outlet stream ($H_2$+$H_2O$) is routed through a condenser to remove steam before measuring hydrogen flow. **b**, Capacity utilization ratio as a function of cycle number. Blue and red markers denote hydrogen

stored and released per cycle, respectively. For Fe-W (solid symbols), the first five redox cycles were intentionally partial to validate the automated operation of the system; subsequent full cycles sustain near-complete utilization (96 ± 3%) with a highly porous microstructure retained in the powder bed. In contrast, a pure Fe powder bed (hollow points) rapidly deactivates, with utilization falling to 11 ± 1% as the powders sinter into a dense mass. **c**, Representative hydrogen release (top) and storage (bottom) process for cycles 10, 20, and 30, demonstrating highly reproducible kinetics. Left axes show instantaneous rate; right axes show cumulative hydrogen stored/released. **d**, Ex-situ XRD patterns of oxidized (top) and reduced (bottom) powder at cycles 10, 20, and 30, confirming reversible phase evolution with highly consistent phase content; broadened $Fe_2W$ reflections in the reduced state indicate the formation of nanocrystalline $Fe_2W$. **e**, Backscattered-electron SEM images of the powder bed in oxidized (top) and reduced (bottom) states for cycles 10, 20, and 30, including loose-powder for surface morphology and cross-sectional views. Phase labels: $FeWO_4$ (FWO, yellow), $Fe_3O_4$ (FO, green), $Fe_2W$ (FW, blue), and Fe (F, orange). Redox operation induces a self-foaming evolution with porosity increasing with cycle number.

**Sintering inhibition mechanism**

The redox reaction process was analyzed using temperature-resolved in-situ XRD (experimental configuration in Figure 3a). A fully oxidized powder mixture containing $Fe_3O_4$ and $FeWO_4$ was heated to 1100 °C under continuous hydrogen flow, while diffraction patterns were acquired sequentially to track phase evolution as a function of temperature. The resulting temperature-resolved in-situ XRD patterns (cascade plot, Figure 3b) highlight the emergence and disappearance of characteristic diffraction peaks from the relevant phases, with the corresponding peak intensity evolution against temperature shown in Figure 3c.

As seen in Figure 3(b,c), $Fe_3O_4$ is reduced first, with its reflections beginning to diminish at ~370 °C as the body-centered cubic (bcc) Fe signal increases, and the reduction is essentially complete by ~500 °C. Reduction of $FeWO_4$ occurs at higher temperatures, initiating at ~610 °C. As $FeWO_4$ reflections attenuate, bcc-W peaks unexpectedly emerge, indicating a transient metallic W formation. With continued reduction, the bcc-W signal gradually decays, accompanied by a modest decrease in the bcc-Fe intensity, while reflections from the intermetallic $Fe_2W$ appear and persist to the end of the experiment. These observations support a multi-step reduction pathway of $FeWO_4$. $FeWO_4$ can be expressed on an oxide basis as $FeO \cdot WO_3$, consistent with common synthesis routes and oxide-component descriptions[17]. During hydrogen reduction, the FeO is reduced to Fe (often described by a shrinking-core model[18,19]),

$$FeO\ (s) + H_2\ (g) \rightarrow Fe\ (s) + H_2O\ (g) \qquad [2]$$

The reduction of $WO_3$ is widely described via a chemical vapor transport (CVT) mechanism[20], in which steam reacts with $WO_3$ to form a volatile oxyhydroxide $WO_2(OH)_2$:

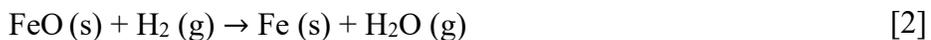

$$WO_3 \text{ (s)} + H_2O \text{ (g)} \rightarrow WO_2(OH)_2 \text{ (g)} \qquad [3]$$

followed by hydrogen reduction to metallic W:

$$WO_2(OH)_2 \text{ (g)} + 3H_2 \text{ (g)} \rightarrow W\text{(s)} + 4H_2O \text{ (g)} \qquad [4]$$

Finally, the transient bcc-Fe and bcc-W react to form the stable $Fe_2W$ intermetallic phase:

$$2Fe \text{ (s)} + W \text{ (s)} \rightarrow Fe_2W \text{ (s)} \qquad [5]$$

A related stepwise reduction sequence has also been reported for $NiWO_4$, where Ni forms first from NiO reduction, followed by progressive $WO_3$ reduction to W[21].

This reaction chemistry underpins two distinct sintering-inhibition mechanisms in the Fe-W system: a dynamic, CVT-enabled self-foaming mechanism that operates during redox cycling, and a static particle-contact-barrier mechanism that retards particle coarsening during high-temperature holds. The dynamic mechanism (Figure 3d) is initiated by transport of volatile $WO_2(OH)_2$, which can redeposit onto previously reduced bcc-Fe surfaces and reduce to bcc-W. The nascent W then locally reacts with Fe to form $Fe_2W$, consistent with the transient bcc-W signature in the in-situ XRD. During the subsequent oxidation step, the stoichiometric mismatch of the Fe-to-W ratio between $Fe_2W$ (Fe: W = 2) and $FeWO_4$ (Fe: W = 1) necessitates partitioning of the excess Fe into an Fe-oxide phase (schematically represented as $Fe_3O_4$ surrounding $FeWO_4$ in Figure 3d), which can become spatially separated from the parent Fe particle. In the next reduction step, the isolated Fe-oxide is reduced back to metallic Fe, whereas $FeWO_4$ again releases more Fe through the FeO → Fe sub-reaction. Over a full redox cycle, repeated volatilization-redeposition thus redistributes both W and Fe, leaving behind nano-voids at the original Fe sites and promoting the nucleation/growth of finer Fe/Fe-oxide grains elsewhere. This cyclic mass transport naturally generates nano-porosity and increasing internal surface area, i.e., a self-foaming evolution, while also favoring nanocrystalline $Fe_2W$ formed from tungsten vapor, consistent with the broadened $Fe_2W$ features observed in the reduced-state XRD (Figure 2d). Such self-foaming behavior resembles the formation of nanoporous metals via vapor-phase dealloying, e.g., via selective removal of Zn as vapor from Zn-Al[22], Zn-Co[23], Zn-Cu[24], or Zn-Mn alloys[25].

When the system is reactively idle at elevated temperature, a second, static mechanism becomes important (Figure 3e). In pure iron beds, oxidation-driven volume expansion promotes particle contact and the resulting internal stresses, together with surface-energy minimization, drive rapid sintering and coarsening of the particle to reduce the total surface area and accommodate volume mismatch; once coarsened in the oxide state, subsequent reduction yields correspondingly larger, sintered metallic Fe particles. Conversely, in Fe-W, the presence of W-containing phases introduces well-distributed barriers between Fe (reduced state) or $Fe_3O_4$ (oxidized state) particles, preventing their merging into larger particles. Together, the dynamic CVT-driven refinement during cycling and the static contact barriers during holds provide a mechanistic basis for the persistence of a porous, kinetically-active microstructure over repeated redox operation.

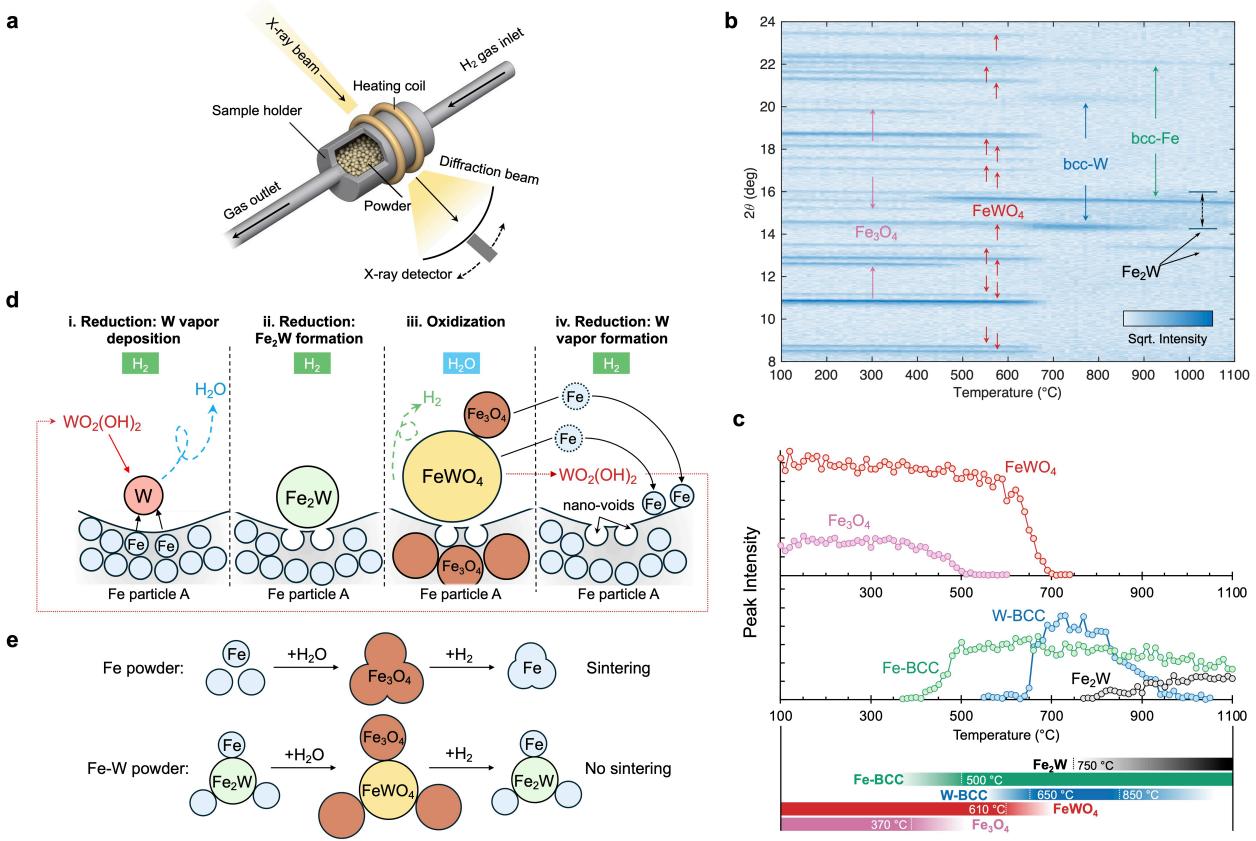

Figure 3: **In-situ XRD reveals the sintering-inhibition mechanisms in Fe-W powders. a**, Schematic of the in-situ XRD configuration. Diffraction patterns were collected while heating the oxidized Fe-W powder ($Fe_3O_4$+$FeWO_4$) to 1100 °C under flowing high-purity $H_2$. **b**, Temperature-resolved in-situ XRD diffraction patterns stacked in a cascade plot, illustrating the sequence of phase transformations during reduction. **c**, Semi-quantitative evolution of phases' integrated peak intensities versus temperature, with onset and completion temperatures for each phase transition. **d**, Proposed CVT-enabled self-foaming mechanism operating during redox cycling, in which volatile $WO_2(OH)_2$ mediates mass transport, generating nano-porosity and progressively refining the microstructure. **e**, Complementary static stabilization mechanism during high-temperature holds, where W-containing phases act as particle-contact barriers and suppress coarsening in both reduced and oxidized states, preventing sintering and preserving powder-bed porosity when redox reactions are inactive.

## Hydrogen storage reaction process

The reduction profile exhibits a two-stage kinetic signature characteristic (Figure 4b): a rapid initial regime (stage 1) transitions to a slower regime (stage 2) after ~40% of the total hydrogen capacity is stored. This inflection coincides with the stoichiometric partitioning of hydrogen consumption in the oxidized bed: $Fe_3O_4 \rightarrow Fe$ accounts for 40.2% of the theoretical uptake, whereas the remaining 59.8% corresponds to the reduction of $FeWO_4$ and formation of $Fe_2W$. This agreement suggests that distinct reactions dominate each kinetic regime.

Both thermodynamic considerations and experimental evidence support this sequential reduction pathway. Temperature-resolved in-situ XRD shows that $Fe_3O_4$ reduces at a lower temperature and earlier than $FeWO_4$ (Figure 3e), indicating a lower kinetic/thermal threshold for $Fe_3O_4$ reduction. Thermodynamic calculations for the $H_2/H_2O$ environment (Figure 4a) further show that $Fe_3O_4$ reduction requires a lower equilibrium hydrogen partial pressure than $FeWO_4$ reduction. Consequently, rapid consumption of $H_2$ during $Fe_3O_4$ reduction can depress the local $H_2$ partial pressure below that required to initiate or sustain $FeWO_4$ reduction, enforcing stage-wise behavior under constant-temperature operation.

We test this hypothesis by interrupting isothermal reduction at ~30% (stage 1) and ~70% (stage 2) capacity and conducting ex-situ XRD (Figure S3b). At ~30% uptake, Fe reduced from $Fe_3O_4$ has formed while $FeWO_4$ remains unreacted, indicating dominance of $Fe_3O_4$ reduction. At ~70% uptake, $Fe_3O_4$ has been fully reduced to Fe, while $Fe_2W$ reflections emerge, confirming that stage 2 mainly corresponds to $FeWO_4$ reduction.

Oxidation does not display an equally sharp two-stage rate transition (Figure 4c), but ex-situ phase analysis indicates an analogous, reversed sequence. When oxidation is interrupted at intermediate extents (30 and 70%) of hydrogen release, the corresponding XRD patterns (Figure S3a) indicate that oxidation is initially dominated by $Fe_2W$ (up to ~60% capacity) and subsequently by Fe at higher extents of hydrogen release.

**Partial-capacity cycling**

Partial-capacity cycling is important from both practical and mechanistic perspectives. In applications requiring rapid charging, the powder bed could be operated at partial conversion, for example, limiting uptake to ~40% of capacity to remain in the fast regime governed primarily by the $Fe \rightleftharpoons Fe_3O_4$ couple. Mechanistically, partial cycling suppresses the participation of the $Fe_2W \rightleftharpoons FeWO_4$ couple and therefore reduces the contribution from the CVT-enabled dynamic self-foaming pathway. Although the static, sluggish-diffusion–mediated sintering resistance remains active, a key question is whether minimizing the dynamic refinement mechanism reintroduces sintering and capacity loss, as reported in previous studies relying on inert sintering barriers [11]. To test this, we designed partial-capacity cycling targeting 25, 50, and 75% of the theoretical capacity, corresponding to none, low, and intermediate involvement of the $Fe_2W \rightleftharpoons FeWO_4$ reaction (and thus the CVT-driven self-foaming), respectively. Each capacity level was run for 30 cycles. Two full-capacity cycles were inserted between partial-cycling sets as diagnostic measurements for the usable capacity. As shown in Figure 4d, these full cycles consistently recovered near-complete capacity utilization after each partial-cycling regime, indicating that the powder bed remains porous and does not exhibit measurable sintering-induced capacity decay under partial-capacity cycling operation.

Microstructural observations support this conclusion (Figure 4f). The morphology of the as-prepared bed and the bed after 0-25% cycling is broadly similar, with no evidence of sintering, suggesting that the static particle-contact-barrier mechanism alone can preserve porosity when CVT-driven foaming is largely inactive. After 0-50% and especially 0-75% cycling, particle refinement becomes increasingly pronounced, consistent with greater engagement of the $Fe_2W \rightleftharpoons FeWO_4$ chemistry and activation of the dynamic self-foaming process. Together, these results indicate that Fe-W can accommodate partial-capacity operation without capacity fade, benefiting from co-existing static and dynamic stabilization mechanisms.

**Temperature-dependent cycling kinetics**

Reaction temperature provides a second lever to balance kinetics and thermal input. Cycling experiments were performed at 650, 700, 750, and 800 °C (examples of storage/release process are shown in Figure S4); below 650 °C, reduction was impractically slow under the present conditions. The temperature dependence of the reaction rate was fitted with an Arrhenius relation (Eqn.[6]),

$$k(T) = A \exp(-\frac{E_a}{RT}) \qquad [6]$$

where k(T) is the rate constant, A is the pre-exponential factor, $E_a$ is the activation energy, R is the universal gas constant, and T is the temperature (Figure 4e).

Hydrogen release (oxidation) is consistently faster than hydrogen storage (reduction) across the studied range. The oxidation point at 800 °C falls below the Arrhenius trend, plausibly due to a steam-feed limitation with the steam supply being held constant across temperatures. Arrhenius fits yield activation energies of 111 kJ mol$^{-1}$ for oxidation and 77 kJ mol$^{-1}$ for reduction, indicating that oxidation is more strongly temperature-activated. This behavior is consistent with oxidation being governed by solid-state processes such as nucleation and growth of oxide and diffusion of oxidants through the product layer, which are highly activated processes[26], whereas reduction can proceed more directly at the metal/oxide interface and often follows a shrinking-core-type progression[18], less hindered by solid-state diffusion. A practical implication of the temperature-dependent study is that thermal efficiency could be improved by operating oxidation at a lower temperature while running reduction at a higher temperature, thereby compensating for the intrinsic kinetic asymmetry between the two half-cycles.

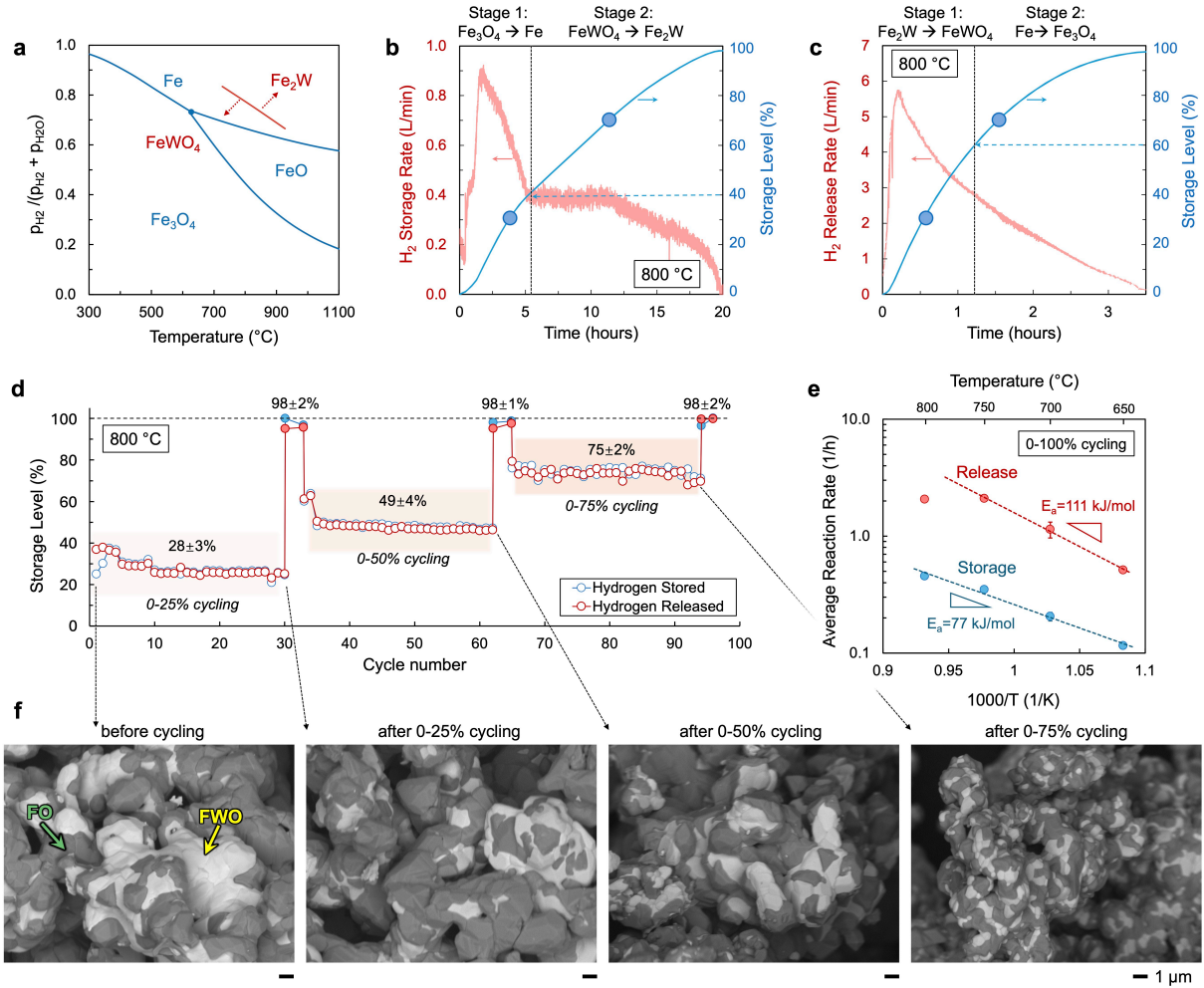

Figure 4: **Reaction pathway, partial-capacity cycling, and temperature-dependent kinetics. a**, Baur-Glässner diagram predicting equilibrium phase as a function of temperature and hydrogen partial pressure (data adapted from ref.[14]). **b**, Representative reduction/storage process (cycle 20 at 800 ºC) highlighting two-stage hydrogen storage kinetics. Stage 1 is dominated by rapid $Fe_3O_3$ reduction, followed by stage 2 governed by slower $FeWO_4$ reduction. Blue markers indicate the ~30% and ~70% storage points at which ex-situ XRD was performed to verify phase dominance in each regime. **c**, Representative oxidation/release process (cycle 20 at 800 ºC). Although the release profile is smooth, ex-situ XRD at ~30% and ~70% release (blue markers) reveals a sequential oxidation pathway in which $Fe_2W$ and Fe oxidize in order. **d**, Partial-capacity cycling designed to evaluate sintering tendency when the CVT-enabled self-foaming mechanism is inactive or only partly engaged. Cycling for 0-25%, 0-50%, and 0-75% capacity was each repeated for 30 cycles, interspersed with two full-capacity cycles (0-100%) to assess recoverable capacity. Near-complete capacity is retained after each partial-capacity cycling test, indicating robust sintering resistance under incomplete operation. **e**, Temperature-dependent kinetics in Arrhenius form. Average reaction rates are shown for reduction (blue) and oxidation (red), with fitted activation energies indicated. The oxidation rate at 800 °C falls below the fitted line, potentially due to the insufficient steam

supply that was kept constant throughout all the control experiments. **f**, Backscattered-electron SEM images of oxidized powders collected across the partial-cycling series. Phase labels: $Fe_3O_4$ (FO, green) and $FeWO_4$ (FWO, yellow). A nearly-unchanged particle size is observed before and after 0-25% cycling, whereas 0-50% and especially 0-75% cycling produce pronounced refinement consistent with engagement of the CVT-enabled self-foaming mechanism.

## Deployment relevance and technology positioning

Benchmarking against incumbent hydrogen storage routes (Figure 1e) places the Fe-W redox platform in a deployment-relevant niche: it emphasizes intrinsic safety, low cost, and high volumetric storage density, while accepting a lower gravimetric energy density that is often non-limiting in stationary settings. The central advance is durability. By suppressing the sintering-driven coarsening that has historically undermined iron-based redox carriers, Fe-W converts an abundant chemistry into a cycle-stable storage medium compatible with repeated operation at high temperature.

This positioning aligns with how hydrogen is used today. Global hydrogen demand remains dominated by established, stationary industrial sectors that occupy >90% of the 2023 market[1] (i.e., petrochemical refining and the chemical industry, notably for ammonia and methanol), so small footprint, safety, and low cost typically outweigh mass-specific metrics for end users. As a result, a compact, solid-phase Fe-W storage unit with robust cycling directly addresses the largest existing demand centers. Beyond stationary deployment, hydrogen/energy global trade via maritime logistics may also value cost and safety over gravimetric performance; reduced Fe-W powders could, in principle, be handled in sealed, moisture-controlled standard containers without pressurized or cryogenic shipboard systems. This market outlook suggests that Fe-W hydrogen storage could capture a share of the global hydrogen storage market projected to reach US$20 billion by 2028 [27]. We anticipate that continued development, particularly reactor-level engineering to improve thermal integration and to enhance reaction kinetics, will accelerate the translation of Fe-W toward an industrially-deployable hydrogen storage technology that supports global net-zero objectives.

## Methods

### Powder preparation

An Fe-25W (at%) precursor blend was produced by mixing $Fe_2O_3$ powder (sourced from Noah Technologies, TX, USA; 99.9wt% purity; particle size < 3 μm) with $WO_3$ powder (sourced from Inframat Advanced Materials, CT, USA; 99.98wt% purity; particle size < 3 μm). The powders were combined in a sealed mixing vessel with zirconia milling media, using a ball-to-powder mass ratio of 1:1 to promote homogenization. To facilitate wet blending and reduce agglomeration, acetone was added as a dispersing solvent (~200 mL per ~300 g of total oxide powder). The suspension was mixed at low-speed rotation (~50 rpm) for 48 h. After mixing, the slurry was transferred to a hot plate and gently heated (~80 °C) for ~48 h to evaporate the acetone. The resulting dried powder was then separated from the zirconia media by sieving, yielding the $Fe_2O_3$-$WO_3$ precursor mixture used in subsequent experiments.

The primary redox experiments shown in Figure 2 and Figure 4(b,c) were conducted using 1.5 kg of the as-prepared $Fe_2O_3$-$WO_3$ precursor powder. To increase experimental throughput, all other tests, including cycling of pure Fe powder, partial-capacity cycling, and temperature-dependent cycling, were performed with 0.150 kg of the same $Fe_2O_3$-$WO_3$ precursor powder. The $Fe_2O_3$-$WO_3$ precursor mixture was poured directly into the container without tapping and used for the following redox-cycling experiments. Following the initial reduction half-cycle at 800 °C, the oxides were converted to a powder bed comprising Fe and $Fe_2W$. Subsequent cycling proceeded reversibly between the reduced equilibrium phase configuration (Fe + $Fe_2W$) and the oxidized configuration ($Fe_3O_4$ + $FeWO_4$).

**Hydrogen storage cycling**

Redox cycling was performed in a custom-built reactor system (Figure 2a). Hydrogen was supplied from an electrolyzer at ~3.7 standard liters per minute (SLM) with a stated purity of 99.9%; for the first 10 cycles in Figure 2b, the hydrogen flow rate was limited to 1.2 SLM by instrument constraints. Steam was generated by metering liquid water directly into the heated zone at a constant rate of 6 mL min$^{-1}$. Inlet and outlet hydrogen flow rates were monitored using mass flow meters (manufactured by Alicat Scientific, AZ, USA).

The Fe-W powder bed was loaded into a cylindrical holder fabricated from perforated 304 stainless steel sheet and mesh (Figure S5; dimensions: 84 mm in diameter, 180 mm in length). Downstream of the reactor, the $H_2$–$H_2O$ effluent was routed through a water-cooled condenser to remove steam by condensation, followed by a desiccant dryer (W. A. Hammond DRIERITE Co. Ltd., OH, USA) to eliminate residual moisture prior to the outlet flow measurement. System operation, including hydrogen generation, water dosing, and synchronized acquisition of flow-meter signals, was automated and logged using a LabVIEW control program.

For the temperature-dependent kinetics analysis, an apparent reaction rate was defined as the reciprocal of the time required to reach 90% of the capacity-equivalent hydrogen storage (reduction) or release (oxidation).

**Microstructural Characterization**

SEM was performed using a JEOL JSM-7900FLV instrument equipped with an energy-dispersive X-ray spectroscopy (EDS) detector. For surface morphology, loose powders were mounted directly onto carbon adhesive tape. For cross-sectional observations, powders were embedded in epoxy, vacuum-infiltrated to ensure pore filling, and then polished to a 1 μm finish.

Powder XRD was carried out on a Rigaku SmartLab diffractometer using Cu Kα radiation. Patterns were collected over 2θ = 10-80° angles at a scan rate of ~1.5 degree/s.

**In-situ XRD Characterization**

In situ XRD measurements were performed using a Stadi-MP diffractometer (Stoe, Germany) equipped with a water-cooled, graphite-heated furnace and Ag Kα radiation (λ = 0.05608 nm). Experiments were conducted under flowing $H_2$ (99.999%, Airgas) at 10 standard cubic centimeters per minute (SCCM) while heating from 30 to 1100 °C at 10 °C/min. The pre-sintered $Fe_3O_4$ + $FeWO_4$ oxide mixture was loaded into a quartz capillary (20 mm length; OD 1.2 mm; ID 1.0 mm) and then inserted into a larger quartz capillary (165 mm length; OD 1.8 mm; ID 1.5 mm). To minimize sample displacement during gas flow and thermal cycling, the capillary assembly was packed with porous ceramic spacers and alumina wool. Diffraction patterns were collected at 10 °C intervals with an exposure time of 60 s per pattern. Phase evolution was quantified by tracking integrated diffraction peak intensities using a custom MATLAB script. The strongest reflections used for integration were bcc-Fe (~15.9°), $Fe_2W$ (~13.4°), $Fe_3O_4$ (~12.7°), $FeWO_4$ (~10.8°), and bcc-W (~14.3°) (2θ, Ag Kα).

**Data availability**

All data supporting the findings of this study are provided in the main text and/or the Supplementary Information. Additional data generated and/or analyzed during the study are available from the corresponding author upon reasonable request.

**Acknowledgments** – The authors gratefully acknowledge the funding and facility support provided by the Northwestern Paula M. Trienens Institute for Sustainability and Energy, the Innovation and New Ventures Office (INVO), and the Trienens-Q Accelerator Program, all at Northwestern University.  They also thank Dr. John Misiaszek (Northwestern University) for useful discussions.

This work made use of: (i) the EPIC facility (RRID: SCR_026361) of Northwestern University's NUANCE Center, which has received support from the IIN and Northwestern's MRSEC program (NSF DMR-2308691); (ii) the MatCI Facility supported by the MRSEC program of the National Science Foundation (DMR-2308691) at the Materials Research Center of Northwestern University; (iii) the IMSERC (RRID:SCR_017874) Crystallography facility at Northwestern University, which has received support from the Soft and Hybrid Nanotechnology Experimental (SHyNE) Resource (NSF ECCS-2025633), and Northwestern University; (iv) the Northwestern University Jerome B. Cohen X-ray Diffraction Core Facility (RRID:SCR_017866) supported by the MRSEC program of the National Science Foundation (DMR-2308691) at the Materials Research Center of Northwestern University and the Soft and Hybrid Nanotechnology Experimental (SHyNE) Resource (NSF ECCS-2025633).

**Author contributions -** Jie Qi: Methodology, Conceptualization, Investigation, Data acquisition and analysis, Visualization, Writing - original draft, Writing – review & editing. David C. Dunand: Methodology, Conceptualization, Funding acquisition, Supervision, Writing - review & editing.

**Declaration of competing interest -** DCD discloses a financial interest in IROX Inc. (USA), which is applying iron-based powders to various redox scenarios. JQ declares no known conflict of interest.

**Declaration of generative AI and AI-assisted technologies in the writing process** - During the preparation of this work, the authors used ChatGPT to improve language clarity and readability of the manuscript. After using this tool, the authors reviewed and edited the content as needed, and they take full responsibility for the content of the publication.